  \documentclass[apj,amsmath,amssymb]{emulateapj-rtx4}

\begin{document}


\title{Models of Kilonova/macronova emission from black hole-neutron star mergers}


\author{Kyohei Kawaguchi}
\affil{Yukawa Institute for Theoretical Physics, Kyoto University, Kyoto 606-8502, Japan}
\author{Koutarou Kyutoku\altaffilmark{1}}
\affil{Interdisciplinary Theoretical Science (iTHES) Research Group, RIKEN, Wako, Saitama 351-0198, Japan}
\author{Masaru Shibata\altaffilmark{1}}
\affil{Yukawa Institute for Theoretical Physics, Kyoto University, Kyoto 606-8502, Japan}
\and
\author{Masaomi Tanaka}
\affil{National Astronomical Observatory of Japan, Mitaka, Tokyo, Japan}
\altaffiltext{1}{Also at Gravitational Wave Research Center, Yukawa Institute for Theoretical Physics, Kyoto University, Kyoto 606-8502, Japan}


\begin{abstract}
 Black hole--neutron star mergers are among the most promising gravitational--wave sources for ground--based detectors, and gravitational waves from black hole--neutron star mergers are expected to be detected in the next few years. Simultaneous detection of electromagnetic counterparts with gravitational wave provides rich information about the merger events. Among the possible electromagnetic counterparts from the black hole--neutron star merger, the emission powered by the decay of radioactive r--process nuclei, so called kilonova/macronova, is one of the best targets for follow--up observation. We derive fitting formulas for the mass and the velocity of ejecta from a generic black hole--neutron star merger based on recently performed numerical relativity simulations. We combined these fitting formulas with a new semi--analytic model for a black hole--neutron star kilonova/macronova lightcurve, which reproduces the results of radiation--transfer simulations. Specifically, the semi--analytic model reproduces the results of each band magnitude obtained by the previous radiation transfer simulations \citep{2014ApJ...780...31T} within $\sim 1~{\rm mag}$.  By using this semi--analytic model, we found that, at $400~{\rm Mpc}$, the kilonova/macronova is as bright as $22$--$24~{\rm mag}$ for the cases with a small chirp mass and a high black hole spin, and $> 28~{\rm mag}$ for a large chirp mass and a low black hole spin. We also apply our model to GRB130603B as an illustration, and show that a black hole--neutron star merger with a rapidly spinning black hole and a large neutron star radius is favored.
\end{abstract}



\section{Introduction}
	The detection of gravitational-waves from a binary--black--hole merger in event GW150914 by Advanced LIGO~\citep{2016PhRvL.116f1102A} marked the start of  gravitational--wave astronomy era. As well as binary--black hole mergers, black hole--neutron star (BH--NS) mergers are among the most promising gravitational-wave sources for ground-based detectors, such as Advanced LIGO, Advanced VIRGO~\citep{2015CQGra..32b4001A}, and KAGRA~\citep{2012CQGra..29l4007S}. The detection of  gravitational waves from BH--NS mergers is expected to be achieved in the next few years because statistical studies predict that more often than once per year of detection can be achieved by those detectors \citep{2010CQGra..27q3001A,2015ApJ...806..263D,2015MNRAS.448..928K}. 
	
	Since a BH--NS binary contains a NS, there are many possible electromagnetic counterparts for the merger event~\citep[e.g.][]{1998ApJ...507L..59L,2011ApJ...736L..21R, 2012ApJ...746...48M,2013MNRAS.430.2585R, 2013MNRAS.430.2121P,2014PhRvD..89f3006T,2013ApJ...775..113T,2014ApJ...780...31T, 2015ApJ...802..119K}. The so--called kilonova/macronova is one of the candidates for the electromagnetic counterparts for a BH--NS merger as well as a NS--NS merger~\citep{1998ApJ...507L..59L}. The kilonova/macronova is induced by the NS material ejected during the merger~\citep{2005ApJ...634.1202R,2013PhRvD..88d1503K, 2015PhRvD..92d4028K,2015PhRvD..92b4014K}. Since the NS consists of highly neutron-rich matter, r-process nucleosynthesis is expected to take place in the ejecta ~\citep{1974ApJ...192L.145L,2010MNRAS.406.2650M}, and the emission powered by decay of the radioactive nuclei would occur. Simultaneous detection of kilonova/macronova and gravitational waves provides rich information about the merger events. It can be useful for determining the host galaxy of the source. As its lightcurve reflects the binary parameters, it could also be useful for extracting the physical information of the binary.
	
	Recently, a variety of numerical-relativity simulations has been performed for BH--NS mergers, and quantitative dependence of the ejecta on a wide range of binary parameters has been revealed~\citep[e.g.][]{2014PhRvD..90b4026F,2015PhRvD..92d4028K,2015PhRvD..92b4014K}. The lightcurve of kilonova/macronova has been also studied by performing radiation transfer simulations by several groups~\citep{2011ApJ...736L..21R,2013ApJ...774...25K,2013ApJ...775..113T,2014ApJ...780...31T}. Specifically,~\cite{2014ApJ...780...31T} performed simulations focusing on BH--NS binaries based on the hydrodynamical evolution obtained by numerical-relativity simulations. These radiation--transfer simulations also consider a detailed heating rate and opacity. However, this study shows the results only for a few ejecta models, and the dependence of the kilonova/macronova emission for a wide range of binary parameters has not been studied yet.
	
	In this paper, we introduce a semi--analytic model for BH--NS kilonova/macronova emission using  fitting formulas for the mass and velocity of dynamical ejecta from BH--NS mergers. We calibrate these fitting formulas by comparing them with the results of recent numerical--relativity simulations for BH--NS mergers performed by Kyoto group. The semi--analytic model for BH--NS kilonova/macronova emission is checked and calibrated by comparing them with the results of a multi--frequency radiation transfer simulation performed by~\cite{2014ApJ...780...31T}. 
	
	 This paper is organized as follows. In Section~\ref{sec:sec2}, we derive fitting formulas for the mass and the velocity of  dynamical ejecta and a semi--analytic model for a kilonova/macronova lightcurve. In Section~\ref{sec:sec3}, we explore the possible range of kilonova/macronova magnitude and its dependence on the binary parameters by using the model derived in Section~\ref{sec:sec2}. In Section~\ref{sec:sec4}, we apply our model to GRB130603B as an illustration. Finally, summaries and remarks of this work are presented in Section~\ref{sec:sec5}. Our convention of notation for physically important quantities is as follows: The BH mass $M_{\rm BH}$, the NS mass $M_{\rm NS}$, the ejecta mass $M_{\rm ej}$, the mass--weighted root mean square of the ejecta velocity $v_{\rm ave}$, the mass ratio $Q=M_{\rm BH}/M_{\rm NS}$, the dimensionless spin parameter of the BH $\chi=cS_{\rm BH}/GM_{\rm BH}^2$, the angle between the BH spin and the orbital angular momentum  $i_{\rm tilt}$, and the compactness of the NS ${\cal C}=GM_{\rm NS}/c^2R_{\rm NS}$. $G$ and $c$ denote the gravitational constant and the speed of light, respectively. The BH and the NS mass are the ADM masses at infinite separation.
	   
\section{Models}\label{sec:sec2}
	In this section we derive fitting formulas for the mass and velocity of dynamical ejecta from the BH-NS mergers, and an analytic model for the lightcurve of a kilonova/macronova. We only consider the dynamical ejecta which is ejected during the merger process, and the ejecta from the BH--accretion disk system, which could be formed in the post--merger phase, is not taken into account. We  remark on the effect of this additional ejecta component in Section~\ref{sec:sec5}.
\subsection{Ejecta Mass}
	\cite{2012PhRvD..86l4007F} introduced a fitting formula for the mass remaining outside the remnant BH after the BH--NS mergers. This remaining mass includes both remnant disk mass and ejecta mass. As the ejecta mass is defined as the gravitationally unbound component of the remaining mass, we expect that the similar form of the fitting model can also be useful for ejecta mass. Therefore, by referring to ~\cite{2012PhRvD..86l4007F}, we propose a fitting model for the ejecta mass as

\begin{eqnarray}
	\frac{M_{\rm ej}}{M_{\rm NS,*}}&=&{\rm Max}\left\{a_1Q^{n_1}\left(1-2\,{\cal C}\right){\cal C}^{-1}\right.\nonumber\\
	&&-a_2\,Q^{n_2}\,{\tilde r}_{\rm ISCO}\left(\chi_{\rm eff}\right)\nonumber\\
	&&+\left.a_3\left(1-\frac{M_{\rm NS}}{M_{\rm NS,*}}\right)+a_4,0\right\},\label{eq:fit}
\end{eqnarray}

\begin{equation}
	\chi_{\rm eff}=\chi\,{\rm cos}\,i_{\rm tilt},
\end{equation}
	and
\begin{eqnarray}
	{\tilde r}_{\rm ISCO}\left(\chi\right)&=&3+Z_2\nonumber\\&-&{\rm sign}\left(\chi\right)\sqrt{\left(3-Z_1\right)\left(3+Z_1+2Z_2\right)},\nonumber\\
	Z_1&=&1+\left(1-\chi^2\right)^{1/3}\nonumber\\&\times&\left\{\left(1+\chi\right)^{1/3}+\left(1-\chi\right)^{1/3}\right\},\nonumber\\
	Z_2&=&\sqrt{3\chi^2+Z_1^2}.
\end{eqnarray}
	 We generalize the model of ~\cite{2012PhRvD..86l4007F} as follows: (i) We set exponents of $Q$, i.e., $n_1$, and $n_2$ as fitting parameters. (ii) We add a term proportional to the specific binding energy of the NS, $1-M_{\rm NS}/M_{\rm NS,*}$, where $M_{\rm NS,*}$ is the total baryon mass of the NS.
	 \begin{table}
\caption{The key quantities of a NS with piecewise polytropic EOSs \citep{2009PhRvD..79l4032R} employed in the numerical--relativity simulations.  $R_{1.35}$, $M_{*,1.35}$, and ${\cal C}_{1.35}$ are the radius, the baryon rest mass, and the compactness parameter for the isolated NS with $M_{\rm NS}=1.35M_\odot$, respectively.}
\begin{center}
 \begin{tabular}{l|ccccc|cccc} \hline
 Model &$R_{1.35}[{\rm km}]$ & $M_{*,1.35}[M_\odot]$ & ${\cal C}_{1.35}$  \\ \hline\hline
 APR4 &   11.1 &  1.50 & 0.180\\
 ALF2 & 12.4 &  1.49 & 0.161\\
 H4 &13.6 &  1.47 & 0.147\\
 MS1 & 14.4 & 1.46 & 0.138\\\hline
 \end{tabular}
\end{center}
\label{tb:eoslist}
\end{table}
	 
	 We use the results of recent numerical--relativity simulations performed by Kyoto group ~\citep{2015PhRvD..92d4028K,2015PhRvD..92b4014K} to determine the fitting parameters, which are summarized in Table~\ref{tb:nr-results}. We use the data for the BH-NS mergers with various mass ratio $Q$, BH spin magnitude $\chi$, BH spin orientation $i_{\rm tilt}$, and NS equation of state (EOS). In the simulations, four phenomenological EOSs described in \cite{2009PhRvD..79l4032R} were employed. We list the key quantities of a NS with the EOS models, which are employed in the numerical--relativity simulations, in Table~\ref{tb:eoslist}. Table~\ref{tb:nr-results} lists the new results of numerical relativity simulations. These new simulations and the computations for initial data are performed by the same method as ~\cite{2015PhRvD..92b4014K}. We note that the NS mass is fixed to be $1.35M_\odot$ for all the models. \cite{2013CQGra..30m5004L} shows that a very massive ejecta ($M_{\rm ej}\gtrsim0.2~M_\odot$) can be produced for the case that the BH spin is extremely high ($\chi_{\rm eff}\sim0.97$). However, because \cite{2013CQGra..30m5004L} is, so far, the only study for such a rapidly spinning case and we do not use its data for the calibration, we only apply the fitting formulas for the case that $\chi_{\rm eff}\le0.9$.

	 We determine the fitting parameters from the simulation data using the least squares method. The best--fit values for the parameters were obtained as follows:
\begin{eqnarray}
	a_1&=&4.464\times 10^{-2},\nonumber\\
	a_2&=&2.269\times 10^{-3},\nonumber\\
	a_3&=&2.431,\nonumber\\
	a_4&=&-0.4159,\nonumber\\
	n_1&=&0.2497,\nonumber\\
	n_2&=&1.352
\end{eqnarray}

 \begin{figure}[h]
	\includegraphics[scale=1]{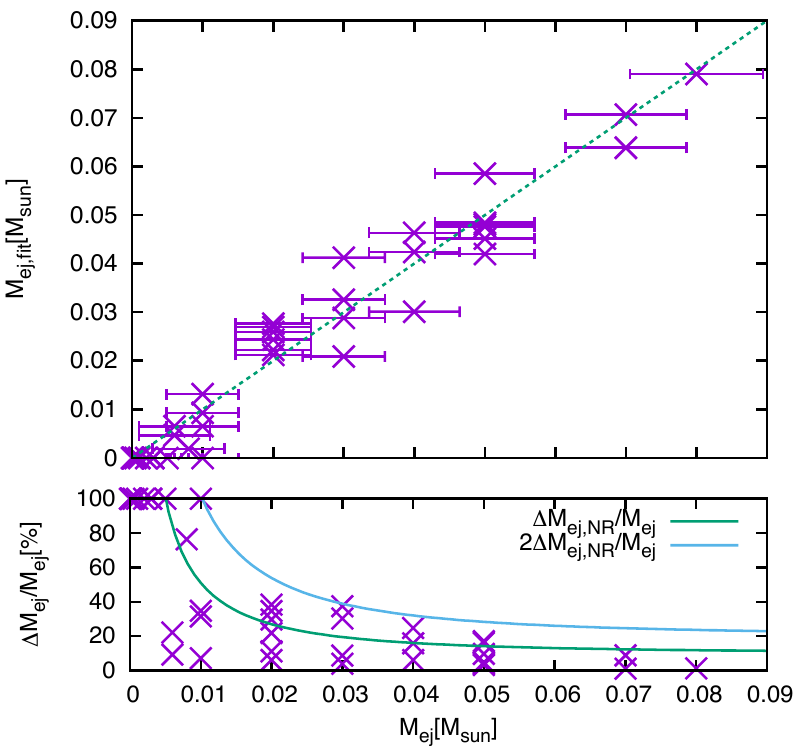}
	\caption{The comparison of the ejecta mass fitting formula with the results of numerical simulations. Each point in the top panel shows the ejecta mass derived by simulations listed in Table~\ref{tb:nr-results} (horizontal axis) and the fitting model using corresponding binary parameters (vertical axis). The errors of the data are estimated using equation (\ref{eq:err}). The bottom panel shows comparison of the estimated relative error of the data and the relative residual error of the ejecta mass fitting model with the best--fit parameters.}
	\label{fig:fitdiff}
\end{figure}

	 Figure~\ref{fig:fitdiff} plots the comparison of the ejecta mass fitting formula with the results of numerical simulations and the relative fitting error of the data as a function of the ejecta mass. The data for $M_{\rm ej}\ge0.05M_\odot$ is fitted within $\sim20\%$, while the data with $0.04M_\odot$ and $0.02M_\odot$ are fitted within $\sim30\%$ and $\sim40\%$, respectively. Because the results obtained by the simulations include errors due to numerical discretization, some dispersion is unavoidable even if the fitting model is appropriate. Since only limited data in Table~\ref{tb:nr-results} were published with explicit error measurement, we assume the estimated numerical error of the simulation data by
\begin{equation}
	\Delta M_{\rm ej,NR}=\sqrt{\left(0.1~M_{\rm ej}\right)^2+\left(0.005 M_\odot\right)^2}\label{eq:err}
\end{equation}
  referring to the estimated numerical error discussed in the Appendix of ~\cite{2015PhRvD..92d4028K} and ~\cite{2015PhRvD..92b4014K}. Figure~\ref{fig:fitdiff} shows that the errors of the fitting are consistent with these estimated errors. The error of the ejecta mass induces the relative error in the peak luminosity only by about a half of its relative error because the luminosity of the kilonova/macronova is approximately proportional to $M_{\rm ej}^{1/2}$ in the early phase. This error in the peak luminosity is comparable to or even smaller than the systematic error of the model of kilonova/macronova lightcurves as we see below. The reduced $\chi^2$ for this fit is defined as,
\begin{equation}
	\chi^2=\frac{1}{N_{\rm NR}-N_{\rm p}-1}\sum^{N_{\rm NR}}_{i=1}\left( \frac{M_{\rm ej}-M_{\rm ej,fit}}{\Delta M_{\rm ej,NR}}\right)^2,
\end{equation}
 where $N_{\rm NR}=45$ and $N_{\rm p}=6$ are the number of the data and the parameter. $\chi^2$ for the model of equation~(\ref{eq:fit}) is $0.85$, and $\chi^2$ become larger than $1$ when we reduced the number of the parameter, such as $a_3$, $a_4$, $n_1$ and $n_2$. Thus, we refrain from increasing the number of parameters in this paper to improve the fitting accuracy, in order to avoid the simulation data to be over--fitted. We note that this fitting formula could have systematic errors due to the choice of the NS EOSs that are used for the fitting. For example, we should check whether our fitting formula can appropriately predict the ejecta mass for two EOSs that give the same NS compactness but give different $M_{\rm NS,*}$ with the same $M_{\rm NS}$. This can only be checked by testing the fitting model with the data using EOS which is not used in this paper. We keep it as a future task.
	 
	We also derive a fitting model for the averaged velocity of the ejecta as a simple linear model of $Q$:
\begin{equation}
	v_{\rm ave}=\left(0.01533\,Q+0.1907\right)~c.
\end{equation}
	The relative error of the fitting is always within $10\%$. However, as is discussed in ~\cite{2015PhRvD..92d4028K}, we expect that the ejecta velocity measured in the numerical--relativity simulation can be overestimated by $\sim 20\%$, and thus, the relative error in the velocity fitting formula can be $\sim 30\%$. This error in ejecta velocity can cause $\sim 15\%$ relative error in $t_{\rm c}$ and the bolometric luminosity in the lightcurve model introduced below, which only weakly affects the following discussions. 
\subsection{Kilonova/Macronova}\label{ssec:2_2}
\begin{figure}[htbp]
	\includegraphics[scale=1.1]{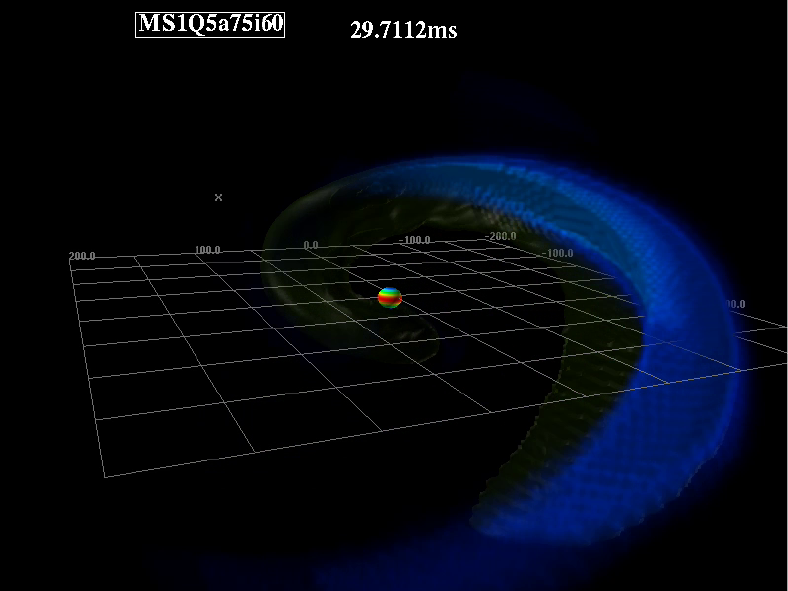}
	\caption{A snapshot of the numerical--relativity result for a BH--NS merger (MS1i60) in \cite{2015PhRvD..92b4014K}. In the figure, the ejecta (unbound material) as well as the location of BH are shown.}
	\label{fig:ejecta}
\end{figure}
\begin{figure*}[htbp]
	\includegraphics[scale=0.45]{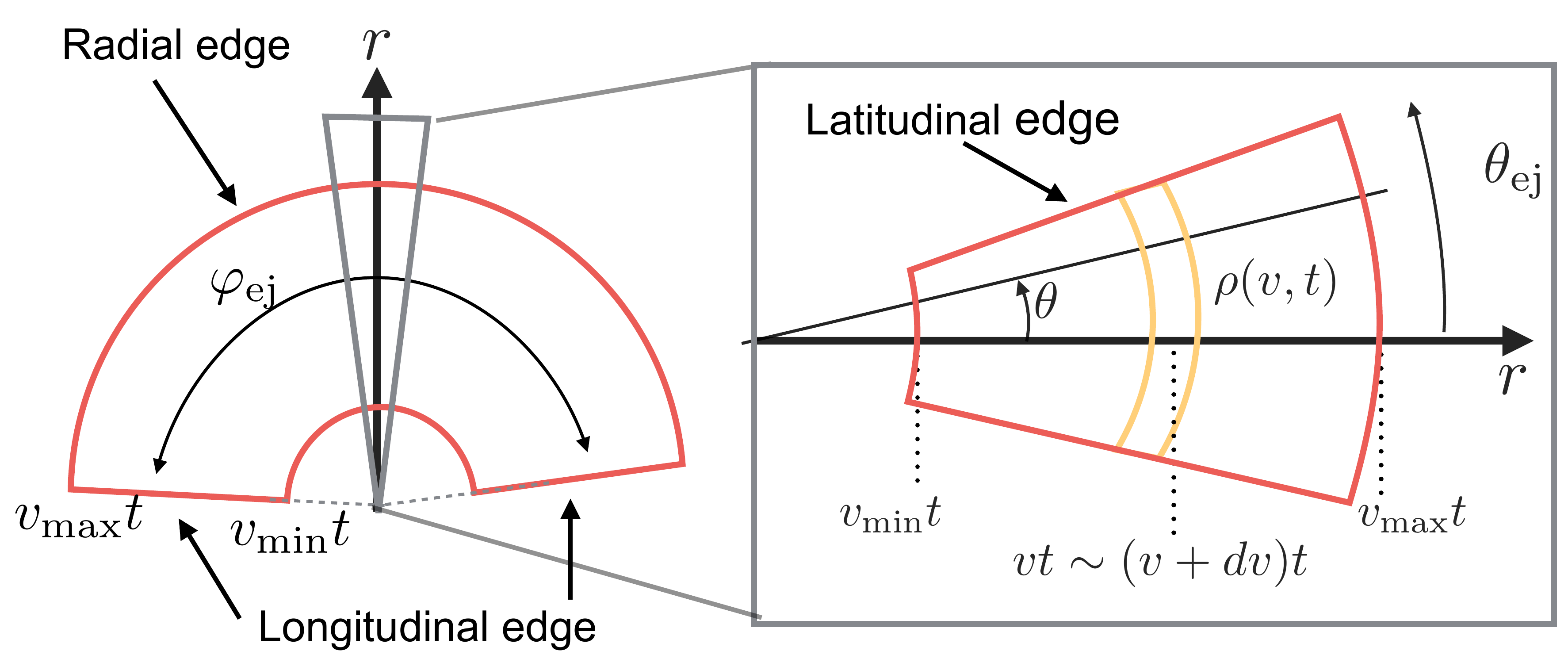}
	\caption{Schematic pictures of the ejecta morphology used in constructing the lightcurve model. $\varphi_{\rm ej}$ and $\theta_{\rm ej}$ are the opening angle and the half thickness of the crescent shape, respectively. $v_{\rm max}$ and $v_{\rm min}$ are the highest and lowest values of the ejecta velocity, respectively.} 
	\label{fig:morph}
\end{figure*}
 Here, we derive a model for the kilonova/macronova from the anisotropic ejecta with velocity distribution resulting from BH-NS mergers in reference to the kilonova/macronova model introduced in ~\cite{2013MNRAS.430.2121P, 2015ApJ...802..119K}. \cite{2013MNRAS.430.2585R,2013PhRvD..88d1503K, 2015PhRvD..92d4028K,2015PhRvD..92b4014K} showed that, the ejecta expands homologously and exhibits a crescent--like shape in most cases for BH-NS mergers (see Figure~\ref{fig:ejecta}). We describe this morphology of the ejecta by modeling the ejecta shape as a partial sphere in the longitudinal and latitudinal directions as shown in Figure~\ref{fig:morph}. We employ spherical coordinates setting the ejecta on the equatorial plane. The latitudinal coordinate $\theta$ is measured from the equatorial plane. Due to the homologous expansion, each shell with the same radius has a velocity $v=r/t$, where $t$ is the elapsed time after the merger.
 
 \cite{2013PhRvD..88d1503K} and ~\cite{2015PhRvD..92d4028K} showed that, the opening angle of its arc is typically $\varphi_{\rm ej}\approx \pi$, while a half thickness of the ejecta in the latitudinal direction is typically $\theta_{\rm ej}\approx 1/5$. The ejecta has approximately flat distribution in its expanding velocity. Assuming a homogeneous mass distribution in the directions of $\theta$ and $\varphi$, and a homologous expansion of the mass shell for each velocity, the density of the ejecta is given by
 \begin{equation}
 	\rho(v,t)=\frac{M_{\rm ej}}{2\varphi_{\rm ej}\theta_{\rm ej}(v_{\rm max}-v_{\rm min})}v^{-2}t^{-3},
 \end{equation}
 where $v_{\rm max}$ and $v_{\rm min}$ are the highest and lowest values of radial velocity of the ejecta, respectively. Here, we assume that ${\rm sin}~\theta_{\rm ej}\approx \theta_{\rm ej}\ll 1$.
 
 We use a diffusion approximation for the radiation transfer. We also assume a gray opacity with $\kappa=10~{\rm cm^2/g}$, which is shown in ~\cite{2013ApJ...774...25K,2013ApJ...775..113T} to be a good approximation for determining the bolometric luminosity of lathanoid--rich kilonovae/macronovae. Because the density of the ejecta decreases in time due to the free expansion, the optical depth of the ejecta decreases with time and hence emission from inner ejecta becomes eventually visible to the observer. Here, in order to calculate the luminosity, we assume that the photons do not diffuse from the radial edge nor the longitudinal edge of the ejecta but only from the latitudinal edge. This assumption is justified by the fact that $\theta_{\rm ej}$ is small for the ejecta from BH-NS mergers and the area of the radial edge and the longitudinal edge are smaller by $\sim \theta_{\rm ej}$ and $\sim \theta_{\rm ej}/\varphi_{\rm ej}$ than the area of the latitudinal edge, respectively. Thus, the contribution of diffused photons to the luminosity is dominated by the photons diffused to the latitudinal direction until the photons starts to escape from the whole ejecta region. This assumption is also consistent with the result of previous radiation transfer simulations, which show that the emission from the radial edge is smaller than the one from the latitudinal edge by a factor of about 3 until $t\sim 1~{\rm day}$~($\sim t_c$ below) (see Figure 3 in~\cite{2011ApJ...736L..21R} or Figure~8 in~\cite{2014ApJ...780...31T}). At the later times, the photons diffuse isotropically since the whole region of the ejecta becomes visible ($t>t_c$). Considering the random walk of photons, the depth of the visible mass is determined by the condition that the distance to the latitudinal edge is comparable to the distance that a photon diffuses, namely $vt(\theta_{\rm ej}-\theta)\approx ct/\tau $. Here,  $\tau\approx\kappa \rho vt (\theta_{\rm ej}-\theta)$ is an optical depth measured from the latitudinal edge. From this condition, we obtain the depth of the visible mass $\theta_{\rm obs}$ as
 \begin{eqnarray}
 	\theta_{\rm obs}(t)&=&\theta_{\rm ej}-\left[\frac{2\varphi_{ej}\theta_{\rm ej}(v_{\rm max}-v_{\rm min}) c}{\kappa M_{\rm ej}}\right]^{1/2}t\nonumber\\
 	&=:&\theta_{\rm ej}\left(1-\frac{t}{ t_{\rm c}}\right),
 \end{eqnarray}
 where we set
 \begin{equation}
 t_{\rm c}=\left[\frac{\theta_{\rm ej}\kappa M_{\rm ej}}{2\varphi_{ej}(v_{\rm max}-v_{\rm min}) c}\right]^{1/2}.
\end{equation}
 
   Using this result, the mass of the photon--escaping region is given by,
 \begin{eqnarray}
 	M_{\rm obs}(t)&=&M_{\rm ej}\nonumber\\
 	&-&\int^{\varphi_{\rm ej}}_{0}d\varphi\int^{\theta_{\rm obs}(t)}_{-\theta_{\rm obs}(t)} {\rm sin}\theta d\theta\int^{v_{\rm max}}_{v_{\rm min}}dv v^2 t^3 \rho \nonumber\\
 	&=&M_{\rm ej}\frac{t}{ t_{\rm c}}~~(t<t_{\rm c}).
 \end{eqnarray} 
 At $t= t_{\rm c}$, the whole region of the ejecta becomes visible. Thus for $t> t_{\rm c}$, we set $M_{\rm obs}(t)=M_{\rm ej}$.
 
  Following ~\cite{2013MNRAS.430.2121P}, we assume that the observed luminosity is dominated by the energy release via radioactive decay. \cite{2012MNRAS.426.1940K, 2014ApJ...789L..39W} showed that, the specific heating rate is given approximately by a power law ${\dot \epsilon}(t)\approx{\dot \epsilon_0}\left(t/{\rm day}\right)^{-\alpha}$, and we set ${\dot \epsilon_0}=1.58\times10^{10}{\rm erg}\,{\rm g}^{-1}~{\rm s}^{-1}$ and $\alpha=1.2$ following~\cite{2014ApJ...780...31T}. The resulting bolometric luminosity is given by
\begin{eqnarray}
	L(t)&=&\left(1+\theta_{\rm ej}\right)\epsilon_{\rm th}{\dot \epsilon_0}M_{\rm ej}\nonumber\\&\times&\left\{
	\begin{array}{ll}
		\displaystyle \frac{t}{ t_{\rm c}}\left(\frac{t}{{\rm day}}\right)^{-\alpha} &t\le  t_{\rm c},\\
		\displaystyle \left(\frac{t}{{\rm day}}\right)^{-\alpha}& t> t_{\rm c}.
	\end{array}
	\right.\label{eq:BL}
\end{eqnarray}
Here, $\epsilon_{\rm th}$ is the efficiency of thermalization introduced in~\cite{2010MNRAS.406.2650M}. The factor $\left(1+\theta_{\rm ej}\right)$ is introduced to include the contribution from the radial edge effectively.

\begin{figure}[htbp]
\includegraphics[scale=1]{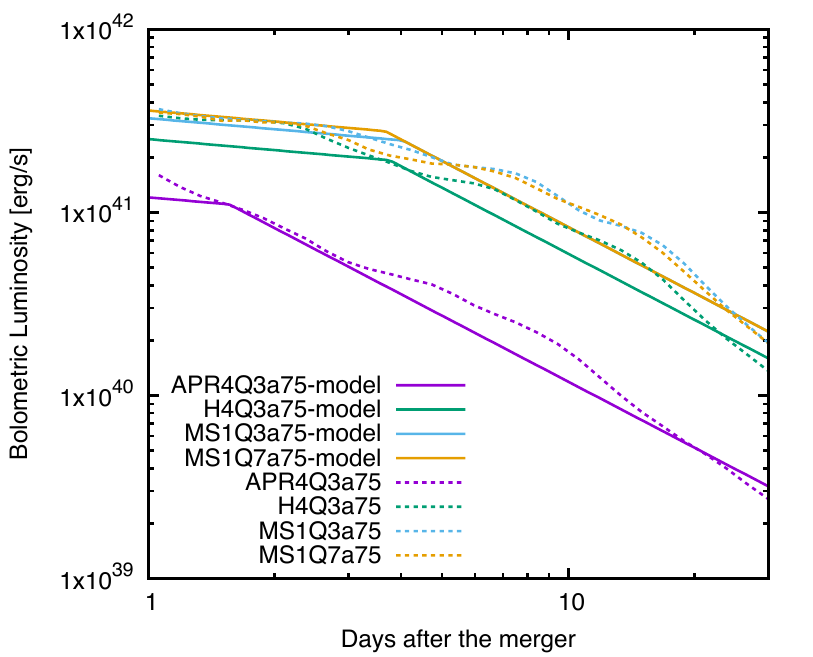}
\caption{Predicted lightcurves of kilonovae/macronovae. The dashed curves denote lightcurves predicted by a radiation transfer simulation performed in~\cite{2014ApJ...780...31T}. The solid curves denote lightcurves predicted by the analytic model obtained in Section~ \ref{ssec:2_2} with the corresponding model parameters. }
\label{fig:th-comp}
\end{figure}

 In order to check the validity of the analytic model obtained above, we compare equation (\ref{eq:BL}) with the results obtained by radiation transfer simulations in ~\cite{2014ApJ...780...31T}\footnote{Data are available at  http:\slash\slash th.nao.ac.jp\slash MEMBER\slash tanaka\slash nsmerger\_lightcurve.html} using the ejecta mass and ejecta velocity profiles obtained by numerical--relativity simulation. Specifically, we focus on the cases for ``APR4Q3a75'', ``H4Q3a75'', and  ``MS1Q3a75'' in ~\cite{2014ApJ...780...31T} and ``MS1Q7a75'' in~\cite{2013ApJ...778L..16H}. We plot the lightcurves predicted by our analytic model in Figure~\ref{fig:th-comp}. Here, we set $\varphi_{\rm ej}=\pi$, $\theta_{\rm ej}=1/5$, $v_{\rm min}=0.02 c$, and $\epsilon_{\rm th}=0.5$ according to ~\cite{2015PhRvD..92d4028K,2012MNRAS.426.1940K}. Using the result of numerical simulations, $(M_{\rm ej}, v_{\rm max})$ are set to be $(1\times 10^{-2} M_\odot,0.41 c)$, $(5\times 10^{-2} M_\odot,0.35 c)$, $(7\times 10^{-2} M_\odot,0.42 c)$, and  $(7\times 10^{-2} M_\odot,0.51 c)$  for ``APR4Q3a75'',  ``H4Q3a75'',  ``MS1Q3a75'', and  ``MS1Q7a75'', respectively. These values of $v_{\rm max}$ are chosen so that mass--weighted root mean squares of the velocity distribution agree with the values of $v_{\rm ch}$ in ~\cite{2014ApJ...780...31T}. As shown in Figure 3, we found that the lightcurves of the analytic model agree with the results of radiation transfer simulations within a factor of $\sim1.4$. 
 
 	The same heating rate and thermalization factor are adopted for both the analytic model and radiation transfer simulation. While the wavelength--dependent opacity is taken into account in the radiation--transfer simulations, we employed gray--opacity with $\kappa=10~{\rm cm^2/g}$ in the analytic model, which is expected to be good approximation for the bolometric lightcurves for lanthanoids--rich ejecta, as shown in \cite{2013ApJ...774...25K,2013ApJ...775..113T}. The much more simplified morphology of the ejecta is used in the analytic model than in~\cite{2014ApJ...780...31T}. Particularly, the parameter of the ejecta morphology, namely $\theta_{\rm ej}$ and $\varphi_{\rm ej}$, are fixed among different binary parameters. \cite{2015PhRvD..92d4028K,2015PhRvD..92b4014K} showed that the ejecta exhibits a similar shape in most cases, particularly if the ejecta mass is larger than $0.01~M_\odot$. At the same time, \cite{2015PhRvD..92d4028K} showed that there is some exception that $\varphi_{\rm ej}$ becomes as large as $2\pi$ with substantial mass $\gtrsim0.01~M_\odot$ (model ALF2-Q7a75). Moreover, $\theta_{\rm ej}$ has a variation up to factor of $\lesssim2$ among the models with substantial mass ejection. The changes of $\theta_{\rm ej}$ and $\varphi_{\rm ej}$ by a factor of 2 change $L$ by $\sim60\%$ and $\sim40\%$, respectively, and $t_c$ by $\sim40\%$. 
 
\subsection{Bolometric Correction}\label{ssec:2_3}
 To predict the magnitude in each wavelength we need to impose an additional assumption because we cannot know the spectra from the analytic model derived in section~\ref{ssec:2_2}. However, the spectra of the kilonova/macronova are determined by very complex frequency-dependent radiative processes of lanthanoids, and hence, it is not easy to model them analytically. Thus, instead, we introduce a phenomenological approach to reproduce the spectra of kilonova/macronova models in the following.

 We define a bolometric correction of a specific band (referred to as $X$--band) $\Delta M_X$ ($X={\it u, g, r, i, z, J, H, K}$) as
\begin{equation}
	\Delta M_X(t):=M_{\rm bol}\left(L(t)\right)-M_X(t),
\end{equation}
	where $M_X$ and $M_{\rm bol}$ are the $X$--band AB magnitude and the bolometric magnitude, respectively. As far as the photospheric emission dominates the total luminosity, the dominant factor to determine the color temperature of the emission, or the bolometric correction, is the temperature near the photosphere. Thus, we need to obtain the temperature at the photosphere to estimate $M_X(t)$. One possible estimator for the temperature at the photosphere is the effective temperature of the surface emission, which is given from equation (\ref{eq:BL}) by the Stefan--Boltzmann law, i.e.,
\begin{eqnarray}
	T_{\rm eff}&=&\left(\frac{L(t)}{\sigma S(t)}\right)^{1/4}\nonumber\\
	&\propto&
	\left\{
	\begin{array}{cc}
		 \left(t/M_{\rm ej}^{1/(2+2\alpha)}\right)^{-1-\alpha}& ~~t\le  t_{\rm c},\\
		 \left(t/M_{\rm ej}^{1/(2+\alpha)}\right)^{-2-\alpha}& ~~t> t_{\rm c},
	\end{array}
	\right.\label{eq:tempeff}
\end{eqnarray}
 where $S(t)\propto t^2$ is the surface area of the latitudinal edge.
 	
	Another simple guess for the temperature at the photosphere is the local temperature in the limit that the radiative cooling is negligible. In this limit, the local internal energy density is proportional to $M_{\rm ej}{\dot \epsilon}(t)t^{-2}$, and assuming that the radiation pressure is dominant there, the local temperature becomes proportional to
\begin{equation}
	T\propto\left(M_{\rm ej}t^{-2-\alpha}\right)^{1/4}=\left(t/M_{\rm ej}^{1/(2+\alpha)}\right)^{-(2+\alpha)/4}.
\end{equation}

	 For both cases, the estimate of the temperature at the photosphere is written as a function $t'=t/M_{\rm ej}^{1/n}$, where $n=2+\alpha$ or $2+2\alpha$.  
	 This suggests that the temperature at the photosphere, and thus, the bolometric corrections are approximately the same for all the models once the time is rescaled by $t'=t/M_{\rm ej}^{1/n}$ with $n=2+\alpha$ or $2+2\alpha$, i.e., $n=3.2$ or $4.4$.

 \begin{figure*}[htbp]
\includegraphics[scale=0.95]{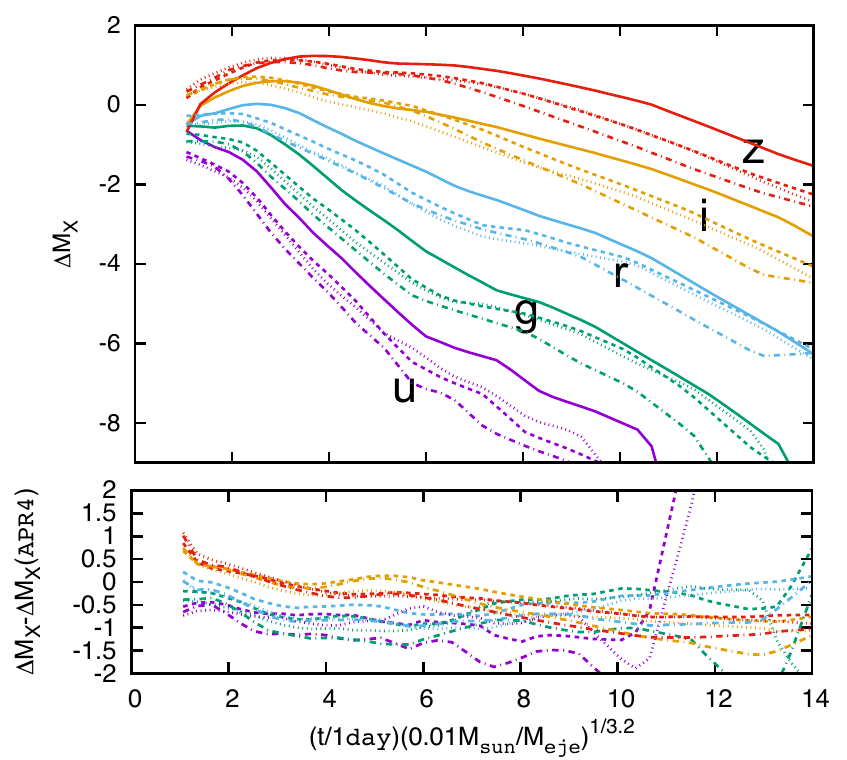}
\includegraphics[scale=0.95]{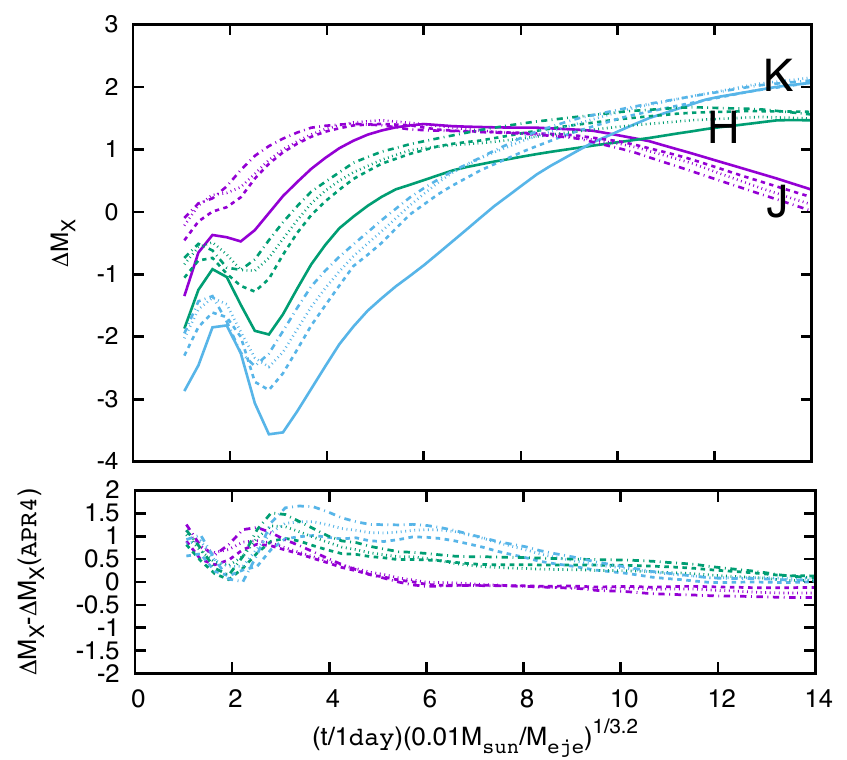}
\caption{The upper panel: the evolution of time-rescaled bolometric corrections calculated from the results from ~\cite{2014ApJ...780...31T}. The solid, dashed, and dotted curves denote the bolometric corrections for APR4Q3a75, H4Q3a75, MS1Q3a75, and MS1Q7a75, respectively. The lower panel: the difference of the time-rescaled bolometric correction of H4Q3a75 (the dashed curves), MS1Q3a75 (the dotted curves), MS1Q7a75 (the dot--dashed curves)  from APR4Q3a75.}
\label{fig:bolcor}
\end{figure*}

	In fact, as is shown in Figure~\ref{fig:bolcor}, we found that the evolution of the bolometric correction of each band filter obtained by radiation transfer simulations of ~\cite{2014ApJ...780...31T} agrees approximately with each other by rescaling the elapsed time by $t/M_{\rm ej}^{1/3.2}$. This result shows that this scaling rule holds in the mass range of $M_{\rm ej}=0.01$--$0.07 M_{\odot}$. This fact allows us to estimate the bolometric corrections for kilonova/macronova with arbitrary ejected mass from the results in each band with a single mass. However, only with a few models with similar binary parameters, for example the spin parameter is fixed to 0.75, the agreement of bolometric correction is tested here. Thus, changing the parameter of the binary can introduce the error on the agreement. The precise effect of binary parameters on the bolometric correction can only be checked by performing the radiation--transfer simulation systematically varying the binary parameters, and, we keep this for the future task.
	
	As we can see from equation (\ref{eq:tempeff}), there is some ambiguity for choosing the value of $n$. Moreover, the emission from the photosphere may not be dominant after the system becomes optically thin. However, time--rescaled bolometric corrections among models with different values of $M_{\rm ej}$ agree approximately with the numerical results in~\cite{2014ApJ...780...31T}, and choosing different value of $n$ leads to only small differences ($<0.5~{\rm mag}$) in the resulting band magnitudes in $t\le 14~{\rm days}$ as long as $n=3.2$--$4.4$. Thus, in this paper, we employ the bolometric correction obtained by rescaling the result of a radiation transfer simulation, specifically the result for APR4Q3a75 in ~\cite{2014ApJ...780...31T}, by employing $n=3.2$ to predict each band magnitude of kilonova/macronova models. The bolometric corrections for {\it ugrizJHK}--band magnitude calculated from the result of APR4Q3a75 in~\cite{2014ApJ...780...31T} are summarized in Table.~\ref{tb:bc}.

\begin{figure*}[htbp]
	\includegraphics[scale=1]{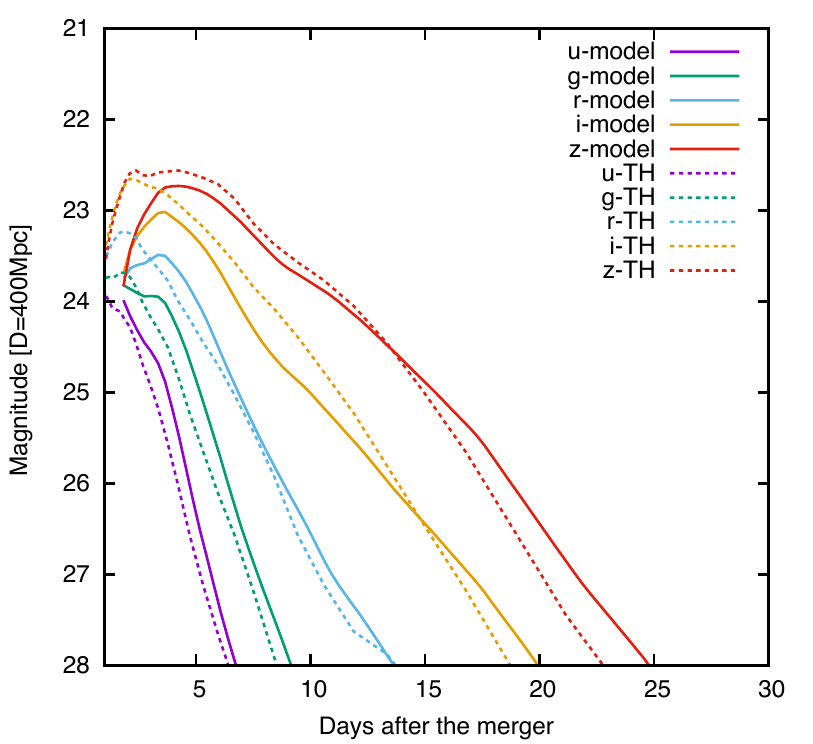}
	\includegraphics[scale=1]{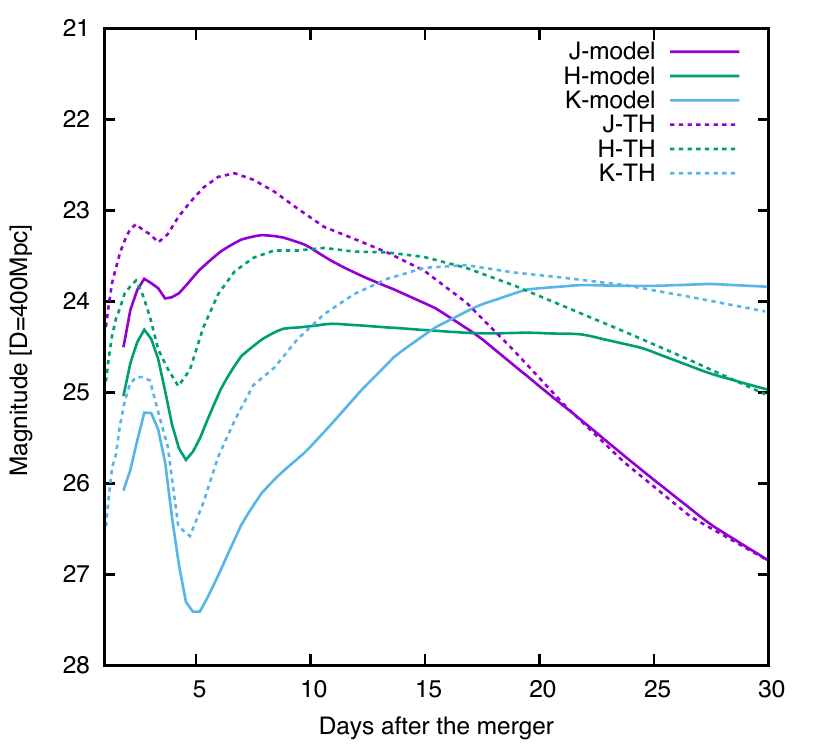}
	\caption{The comparison of {\it ugriz}--band (left figure) and the {\it JHK}--band (right figure) AB magnitudes between the results obtained in ~\cite{2014ApJ...780...31T} (dashed lines) and the light-curve models that are derived from the ejecta mass model and the bolometric correction model (solid lines) for H4Q3a75.}
	\label{fig:th-mag-comp}
\end{figure*}
	Figure~\ref{fig:th-mag-comp} shows a comparison of the {\it ugrizJHK}--band magnitudes between the model obtained in Section~\ref{ssec:2_2} and the results of radiation transfer simulations in~\cite{2014ApJ...780...31T}. The band magnitudes are calculated with the combination of the mass and the velocity fitting formulas, the bolometric luminosity model, and the bolometric correction model (we call this model ``the analytic model'' in the following). We set a hypothetical distance as $400\,{\rm Mpc}$, which is the typical effective distance to a BH--NS merger event expected to be detected by the gravitational--wave detectors \citep{2010CQGra..27q3001A}.  For each band magnitude, we find that the analytic model always agrees with the previous results of radiation transfer simulations within $\sim 1~{\rm mag}$\footnote{Online lightcurve calculator of our model is available at  http:\slash\slash www2.yukawa.kyoto-u.ac.jp\slash \~{}kyohei.kawaguchi\slash kn\_calc\slash main.html}.  We note that the variation in the ejecta geometry and the uncertainties in the opacity and the heating rate can be the sources of errors in the lightcurve model and the radiation transfer simulations as we discuss in section~\ref{sec:sec5}.

\section{Implications for Observing Strategies}\label{sec:sec3}
	Once the detection of gravitational waves from a compact binary merger is achieved, the chirp mass ${\cal M}_{\rm chirp}=[Q^3/(1+Q)]^{1/5}M_{\rm NS}$, the symmetric mass ratio $\nu=Q/(1+Q)^2$, and the effective spin parameter $\chi_{\rm eff}=\chi\,{\rm cos}\,i_{\rm tilt}$ of the binary will be estimated ~\citep[e.g.,][]{2014PhRvD..89f4048O,2014PhRvD..89j2005O}. Assuming a BH--NS merger event, the analytic model we constructed in this paper predicts the brightness and the duration of the kilonova/macronova for detected events. 
	~\cite{2014ApJ...780...31T} pointed out that the observation in {\it i}--band filter by wide--field 8--m class telescopes, such as Subaru/Hyper Spurime Cam \citep[HSC;][]{2006SPIE.6269E..0BM} and Large Synoptic Survey Telescope \citep[LSST;][]{2008arXiv0805.2366I}, will be useful for the follow--up observations of kilonovae/macronovae. It is also pointed out that near--infrared observations by wide--field space telescopes, such as The Wide-Field Infrared Survey Telescope \citep[WFIRST;][]{2012arXiv1208.4012G}, will be promising. Thus, for an illustration, we show the evolutions of {\it i} and {\it H}--band magnitudes and  parameter dependence of BH--NS kilonovae/macronovae in this section.

\begin{figure}[h]
	\includegraphics[scale=1]{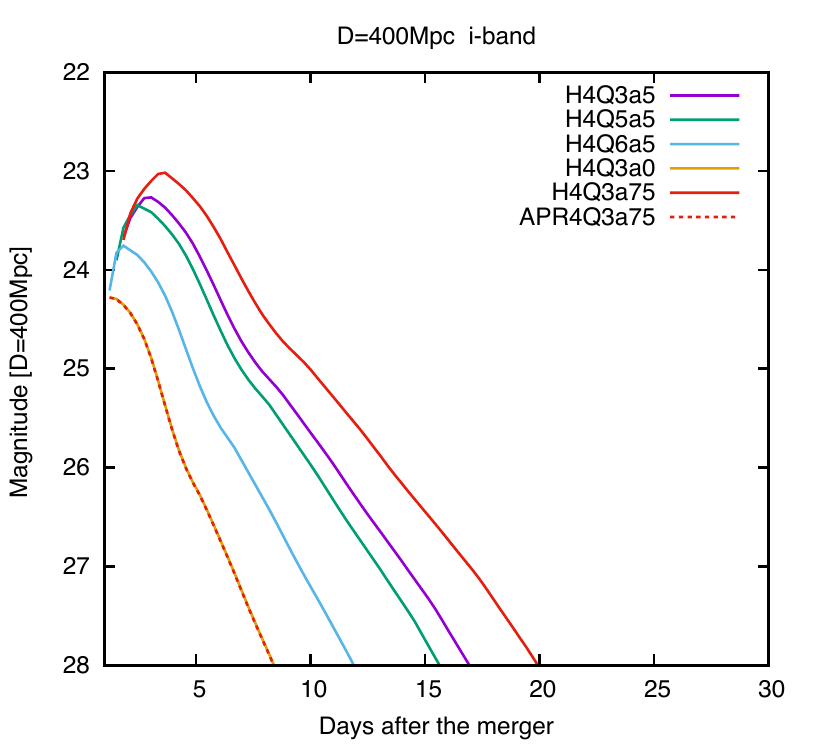}
	\caption{The {\it i}--band AB magnitude of kilonovae/macronovae calculated by the analytic model for various binary parameters. The label for the model denotes the EOS name, the mass ratio $Q$, and the effective spin parameter $\chi_{\rm eff}$ of the binary. Specifically,  ``Q3'', ``Q5'', and ``Q6'' denote the models with $Q=3$, $5$, and $6$, respectively. ``a0'', ``a5'', and ``a75'' denote the models with $\chi_{\rm eff}=0$, $0.5$, and $0.75$, respectively. We set $M_{\rm NS}=1.35M_\odot$, and the distance to be $400\,{\rm Mpc}$.}
	\label{fig:magdep}
\end{figure}
\begin{figure*}
	\includegraphics[scale=0.95]{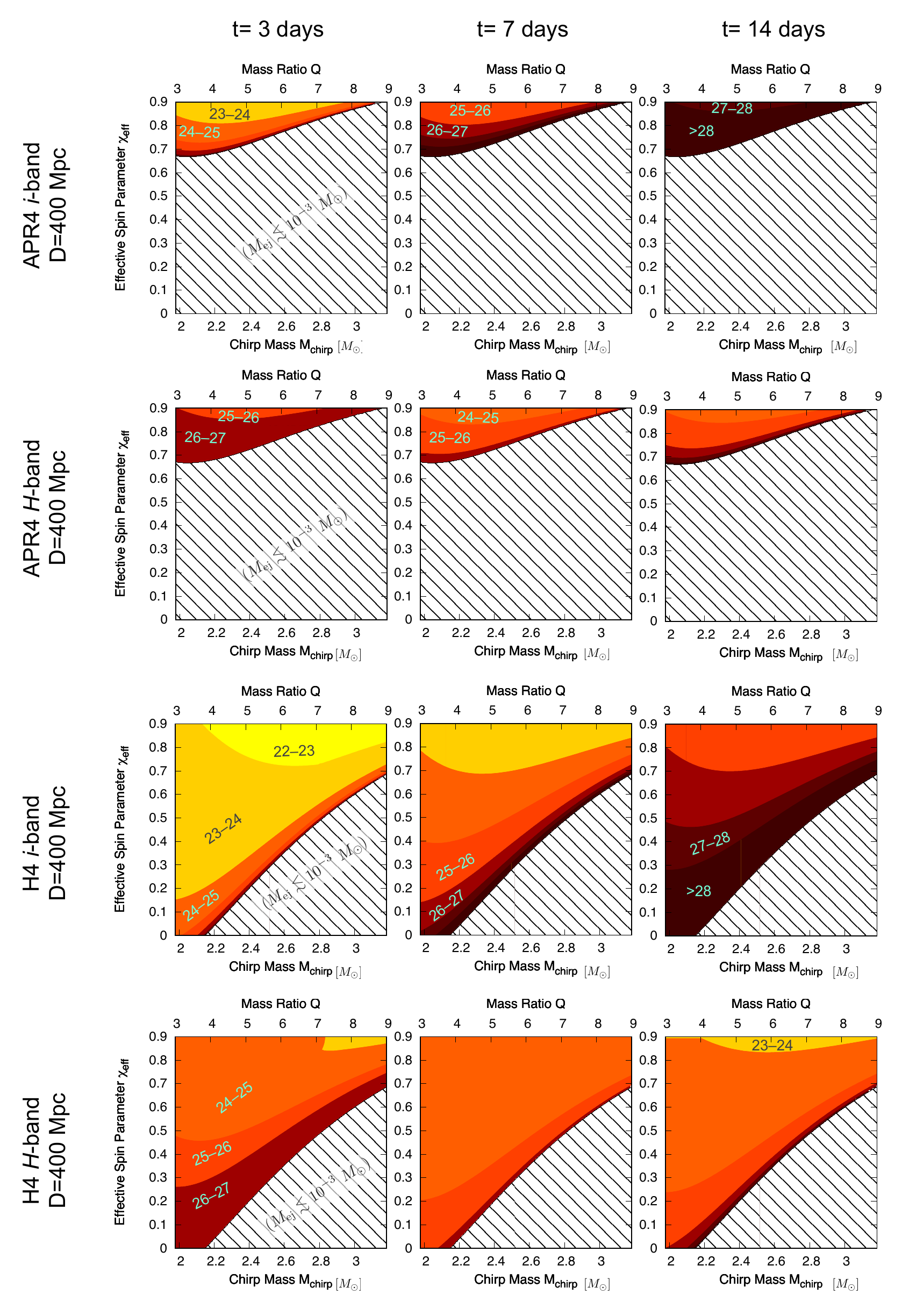}
	\caption{Expected {\it i} and {\it H}--band AB magnitudes as functions of the chirp mass ${\cal M}_{\rm ch}$ (lower horizontal axis) or the mass ratio $Q$ (upper horizontal axis) and the effective dimensionless spin parameter $\chi_{\rm eff}$. The mass of the NS is set to be $M_{\rm NS}=1.35M_\odot$ and the distance to the BH--NS binary is set to be $400 ~{\rm Mpc}$.}
	\label{fig:mag-cont}
\end{figure*}

	In Figure~\ref{fig:magdep}, we plot the evolution of {\it i}--band magnitude for several binary parameters. The label for the model denotes the EOS name, the mass ratio $Q$, and the effective spin parameter $\chi_{\rm eff}$ of the binary. Specifically, ``Q3'', ``Q5'', and ``Q6'' denote the models with $Q=3$, $5$, and $6$, respectively. ``a0'', ``a5'', and ``a75'' denote the models with $\chi_{\rm eff}=0$, $0.5$, and $0.75$, respectively. We set $M_{\rm NS}=1.35M_\odot$. We can see that the kilonova/macronova model becomes brighter as $Q$ gets smaller and $\chi_{\rm eff}$ gets larger. This dependence reflects the fact that the ejecta mass increases with the decrease of the mass ratio of the binary and with the increase of the effective spin of the BH. We note that, exceptionally, the dependence of the luminosity on the mass ratio is not monotonic for rapidly spinning case $\chi_{\rm eff}\gtrsim 0.75$, as seen in Figure~\ref{fig:mag-cont}. The kilonova/macronova model is always dimmer for APR4 than for H4, which also reflects the fact that the NS with a larger radius produces more massive ejecta. This dependence of the ejecta mass has been shown by previous numerical simulations ~\citep[e.g.,][]{2015PhRvD..92d4028K,2015PhRvD..92b4014K}.

	In Figure~\ref{fig:mag-cont}, we show the contour plots of {\it i} and {\it H}--band magnitudes at $t=3, 7,$ and $ 14$ days after the merger for APR4 and H4 as a function of chirp mass ${\cal M}_{\rm chirp}=[Q^3/(1+Q)]^{1/5}M_{\rm NS}$ and $\chi_{\rm eff}=\chi\,{\rm cos}\,i_{\rm tilt}$. The magnitudes and the chirp mass are calculated assuming $M_{\rm NS}=1.35M_\odot$. The distance is set to be $400{\rm Mpc}$. We cut the regions from the plot in which the ejecta mass is below $10^{-3}~M_\odot$ because the data of the radiation--transfer simulation is not long enough to calculate the bolometric correction at $14~{\rm days}$ after the merger for the case that the ejecta mass is below $10^{-3}~M_\odot$. We note that {\it i} and {\it H}--band magnitude are always dimmer than $26~{\rm mag}$ in those regions for both APR4 and H4.  

	For the case that the NS EOS is H4, {\it i}--band magnitude can reach $\sim 23~{\rm mag}$ for rapidly spinning BH, $\chi_{\rm eff}\ge0.5$ for a wide range of $Q$. The emission in {\it i}--band with $\le 26~{\rm mag}$ can last for $\sim7$ days after the merger. On the other hand, a BH--NS merger with small BH spin and large chirp mass, particularly with $\chi_{\rm eff}\lesssim0.67~({\cal M}_{\rm ch}/M_\odot)-1.45$, produces no ejecta or ejecta with mass below $10^{-3}~M_\odot$ and leads to a kilonova/macronova dimmer than $\sim 26~{\rm mag}$ in {\it i}--band. In contrast to the {\it i}--band magnitude, the {\it H}--band magnitude becomes brighter as the time elapses. This time dependence reflects the fact that the color temperature decreases with time, and thus, the bolometric correction in {\it H}--band increases with time. Due to this time dependence, the {\it H}--band magnitude is brighter than $25~{\rm mag}$ until $\sim14~{\rm days}$ if the BH has sufficiently large spin parameter: If $\chi_{\rm eff}\gtrsim0.46~({\cal M}_{\rm ch}/M_\odot)-0.72$.
	
	For the case that the NS EOS is APR4, the ejecta heavier than $10^{-3}~M_\odot$ is produced when $\chi_{\rm eff}\gtrsim0.23~({\cal M}_{\rm ch}/M_\odot)-0.18$. There is no region where the {\it i}--band magnitude reaches $ 23~{\rm mag}$ as long as $\chi_{\rm eff}\le0.9$, and only a very narrow region with a very high BH spin ($\chi_{\rm eff}\gtrsim 0.8$) can reach $24~{\rm mag}$ only until the first $\sim3$ days. At $7~{\rm days}$ after the merger, the {\it i}--band magnitude is dimmer than $26~{\rm mag}$ in the most of the region shown in the plot. Exceptionally, the {\it i}--band magnitude reaches $26~{\rm mag}$ for the case that $\chi_{\rm eff}\gtrsim 0.8$ and ${\cal M}_{\rm ch}\lesssim 2.8~M_\odot$. Similarly, the {\it H}--band magnitude reaches $25~{\rm mag}$ $\sim7$ days after the merger for the case that $\chi_{\rm eff}\gtrsim 0.85$ and ${\cal M}_{\rm ch}\lesssim 2.8~M_\odot$, and always dimmer than $25~{\rm mag}$ for other cases.
	
Our results provide a guide for electromagnetic follow-up observations. In optical wavelengths, 
in order to maximize the possibility to detect electromagnetic counterparts,
follow-up observations with 8--m class telescopes 
within $\sim 3~{\rm days}$ after the merger are crucial.
A typical limiting magnitude for the 8--m class telescopes
is $\sim 26~{\rm mag}$ for $10~{\rm min}$ exposure.
Therefore, rapid follow-up observations with 8--m class telescopes will
open the possibility to detect the kilonova/macronova emission
even for the faint models with soft EOS (APR4).

In near-infrared wavelengths,
the emission brighter than $25~{\rm mag}$ lasts for $t\sim 14~{\rm days}$.
Since a limiting magnitude for the infrared telescopes such as WFIRST
is $\sim 25~{\rm mag}$, infrared satellite observations will give us more chances
to find the kilonovae/macronovae from the BH--NS merger.
	
\section{Application to GRB130603B}\label{sec:sec4}
	The analytic model we introduced in Section~\ref{sec:sec2} can also be used to constrain the binary parameters by kilonova/macronova emissions. GRB130603B is a gamma--ray burst which is possibly associated with kilonova/macronova ~\citep{2013Natur.500..547T,2013ApJ...774L..23B}. \cite{2015NatCo...6E7323Y,2015ApJ...811L..22J} show that GRB060614 could also be a gamma--ray burst with a kilonova/macronova association. We apply our model to a possible kilonova/macronova candidate associated with GRB130603B in this section. 
	
	A near-infrared excess from an afterglow model associated with GRB130603B was observed by the Hubble Space Telescope (HST) in {\it H}--band ~\citep{2013Natur.500..547T,2013ApJ...774L..23B}. According to their discussion, the redshift of the host galaxy is $z=0.356$, and the near-infrared emission corresponds to an emission with {\it J}--band absolute magnitude $M_{{\it J},{\rm abs}}=-15.7~{\rm mag}$ at $\approx 7\,{\rm days}$ after the prompt emission in the gamma--ray burst rest frame. 
\begin{figure*}
	\includegraphics[scale=1.6]{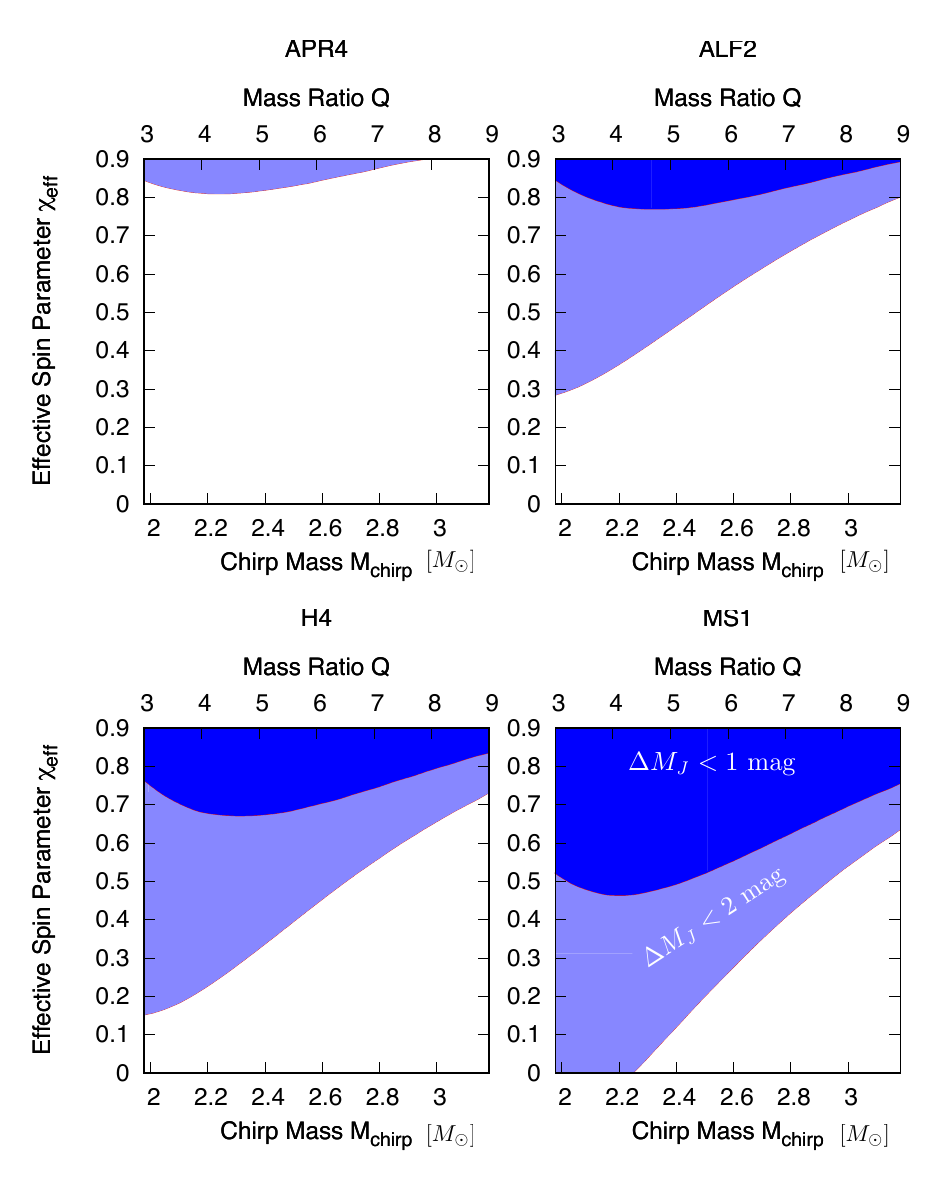}
	\caption{Allowed parameter region for explaining the observation of GRB130603B. The results for four EOSs are shown. The blue and light--blue colored regions show the parameter space in which the difference of {\it J}--band magnitude between the prediction and the observation of GRB130603B obtained by HST is within 1 mag and 2 mag, respectively.}
	\label{fig:grb-cont}
\end{figure*}

	We calculate {\it J}--band magnitudes at $7~{\rm days}$ after the merger by using the analytic model, $M_{{\it J},{\rm model}}$, and plot the difference between those magnitudes and the observed magnitude $\Delta M_{\it J}=M_{{\it J},{\rm model}}-M_{{\it J},{\rm abs}}$ as a function of ${\cal M}_{\rm chirp}$ and $\chi_{\rm eff}$ in Figure~\ref{fig:grb-cont}. The magnitudes and the chirp mass are calculated assuming $M_{\rm NS}=1.35M_\odot$. Four different cases employing APR4, ALF2, H4, and MS1 EOS are shown. The magnitude obtained by the analytic model is always larger than the observed magnitude. We note that our models tend to be fainter than the results of radiative
transfer simulations as we see Figure~\ref{fig:th-mag-comp}, and thus, this comparison may be conservative.
	
	  The blue and light--blue regions in Figure~\ref{fig:grb-cont} show the regions where the difference of {\it J}--band magnitude between the prediction and the observation is within 1 mag and 2 mag, respectively. Supposing that the error of the analytic model is within $1~{\rm mag}$, the binary parameters are constrained to this blue region. We note that the observational errors or uncertainties are smaller than the uncertainties in the model. The observational error of the HST {\it H}--band data is about $0.3~{\rm mag}$~\citep{2013Natur.500..547T}. The Galactic extinction in the direction to GRB130603B is negligible ($A_{\it H}\sim0.01~{\rm mag}$). The extinction in the host galaxy is estimated to be $A_{\it V}\sim0.8$--$1.0 ~{\rm mag}$~\citep{2014A&A...563A..62D,2014ApJ...780..118F}, and thus, according to the Galactic extinction law ($A_{\it J}/A_{\it V}~\sim0.282$,~\cite{1989ApJ...345..245C}), the extinction in ${\it J}$--band can be estimated as $A_{\it J}\approx0.23$--$0.28~{\rm mag}$. Figure~\ref{fig:grb-cont} shows that a BH--NS merger with small spin $\chi_{\rm eff}\le0.3$ is disfavored for all EOS cases. In particular, for the case that the NS EOS is APR4, there is no region with $\chi_{\rm eff}\le 0.9$ in which the predicted magnitude is consistent with the observation within $1~{\rm mag}$. Thus, the NS with a small radius is disfavored from this observation (unless the BH is rapidly spinning with $\chi_{\rm eff}>0.9$) if the ejecta is dominated by dynamical ejecta, which is also pointed out by  ~\cite{2013ApJ...778L..16H} using the models in a smaller parameter region. Applying the same arguments to GRB060614, we obtain similar results, and BH--NS mergers with a large NS radius, a small mass ratio, or a rapidly spinning BH are favored, which is consistent with the discussion in ~\cite{2015NatCo...6E7323Y}.

\section{Summary \& Remarks}\label{sec:sec5}
	We developed fitting formulas for the mass and the velocity of dynamical ejecta from a generic BH--NS merger based on recently--performed numerical relativity simulations. We combined these fitting formulas with a semi-analytic model for the BH--NS kilonova/macronova lightcurve that reproduces the results of radiation transfer simulations with a small error. Specifically, the semi--analytic model reproduces the results of each band magnitude obtained by previous radiation transfer simulations within $\sim 1~{\rm mag}$.
	 
	 This semi--analytic model shows that the kilonova/macronova can be observed in optical wavelength by 8--m class telescopes within $3~{\rm days}$ from the merger if $\chi_{\rm eff}\gtrsim0.23~({\cal M}_{\rm ch}/M_\odot)-0.18$ for the case that the NS EOS is stiffer than APR4, namely for the case that the NS radius is larger than $\sim11~{\rm km}$. On the other hand, if the optical--wavelength emission is observed by 8--m class telescopes for the case that  $\chi_{\rm eff}\lesssim0.67~({\cal M}_{\rm ch}/M_\odot)-1.45$, the origin of the emission may not be the kilonova/macronova from the dynamical ejecta but some other components unless the NS EOS is stiffer than H4. In near--infrared wavelengths, the follow--up observation of kilonova/macronova by space telescopes is useful because the emission can be bright and last long enough to be detected for the wide parameter space of a BH--NS merger. Particularly, if the NS EOS is stiff as H4, the emission can be detected even at $\sim14~{\rm days}$ after the merger for the case that $\chi_{\rm eff}\gtrsim0.46~({\cal M}_{\rm ch}/M_\odot)-0.72$. We also applied our model to GRB130603B as an illustration, and showed that a BH--NS merger with a rapidly spinning BH or a large NS radius is favored, which is consistent with the previous studies \citep{2013ApJ...778L..16H}.

	There are several error sources for the results of the analytic model and the radiation transfer simulations. The variation in the ejecta geometry can be the source of the error in the lightcurve model. It can change the lightcurve of the analytic model by $\sim 0.5~{\rm mag}$. The errors in the mass and the velocity fitting formulas can also induce the errors in the lightcurves by $\sim 0.6~{\rm mag}$.  Moreover, the uncertainties of the opacity and the heating rate can be additional factors which may affect the results. Since there are still some uncertainties in the opacity for lanthanoids, the real opacity can change from the one which is employed in both the analytic model and the simulations in~\cite{2014ApJ...780...31T}.  For example, because the bolometric luminosity is proportional to $\kappa^{-1/2}$ in $t<t_{\rm c}$, the increase of $\kappa$ by a factor of $3$ makes the bolometric magnitude large by $0.6~{\rm mag}$. The uncertainties in the heating rate and the thermal efficiency can induce the error in the lightcurve linearly.

	We should also note that there may exist some other ejecta components which are produced in the post--merger phase and can be an additional source for the emission. For example, ~\cite{2015PhRvD..92f4034K}, \cite{2015MNRAS.448..541J}, and \cite{2015MNRAS.446..750F} show that mass comparable to (or even more than) the dynamical ejecta may be ejected as a disk wind from a BH--NS merger remnant. While the screening effect of subsequent ejecta by preceding ejecta is argued for NS--NS mergers ~\citep{2015MNRAS.450.1777K},  this may not take place for the kilonova/macronova from BH--NS mergers because the dynamical ejecta is confined in a narrow region around the equatorial plane: Additional components of ejecta could enhance the electromagnetic emission. Thus, additional components may be visible and the emission can be brighter than we predict in this paper. This may make it difficult to constrain the binary parameters from kilonova/macronova observations. For example, additional components may allow a larger parameter region which is consistent with GRB130603B to exist even for the case that the NS EOS is as soft as APR4. Therefore, we should take possibilities of other emission into account to constrain the models, and we keep this as our future task.
	
	We thank to F. Foucart for helpful comments on the preprint version of the manuscript. This study was supported in part by the Grants-in-Aid for the
Scientific Research of JSPS (14J02950, 24244028, 24740117, 15H02075, 15H06857), MEXT (25103515, 15H00788), and RIKEN iTHES project.

\begin{table*}
\begin{center}
	\caption{The list of results obtained by recent numerical--relativity simulations: $Q$, $\chi$, $i_{\rm tilt}$, $M_{\rm ej}$, and $v_{\rm ave}$ are the mass ratio of the binary, the dimensionless spin parameter of the BH, the initial misalignment angle between the BH spin and the orbital angular momentum, the ejected mass, and the mass--weighted root mean square of the ejecta velocity distribution, respectively. ``New'' in the ``Ref'' column denotes the data obtained by our new numerical simulations.}
	\label{tb:nr-results}
	\begin{tabular}{c|cccc|ccc}\hline
		ID	&	$Q$	&	$\chi$	&	$i_{\rm tilt}$~[$^\circ$]	&	EOS		&	$M_{\rm ej}$	&	$v_{\rm ave}$~[$c$]		&	Ref\\\hline\hline
		1	&	3	&	0	&	0	&	APR4	&	$<1\times 10^{-3}$	&	0.20	&	[1]	\\
		2	&	3	&	0	&	0	&	ALF2	&	0.003	&	0.22	&	[1]	\\
		3	&	3	&	0	&	0	&	H4		&	0.006	&	0.22	&	[1]	\\
		4	&	3	&	0	&	0	&	MS1		&	0.02 	&	0.23	&	[1]	\\
		5	&	3	&	0.5	&	0	&	APR4	&	0.002	&	0.21	&	[1]	\\
		6	&	3	&	0.5	&	0	&	ALF2	&	0.02 	&	0.24	&	[1]	\\
		7	&	3	&	0.5	&	0	&	H4		&	0.03 	&	0.23	&	[1]	\\
		8	&	3	&	0.5	&	0	&	MS1		&	0.05 	&	0.24	&	[1]	\\
		9	&	3	&	0.75	&	0	&	APR4	&	0.01 	&	0.23	&	[1]	\\
		10	&	3	&	0.75	&	0	&	ALF2	&	0.05 	&	0.25	&	[1]	\\
		11	&	3	&	0.75	&	0	&	H4		&	0.05 	&	0.24	&	[1]	\\
		12	&	3	&	0.75	&	0	&	MS1		&	0.07 	&	0.25	&	[1]	\\
		13	&	3	&	0.75	&	31	&	H4		&	0.03 	&	0.22	&	New	\\
		14	&	3	&	0.75	&	62	&	H4		&	0.02 	&	0.24	&	New	\\
		15	&	3	&	0.75	&	93	&	APR4	&	$<1\times 10^{-3}$	&	0.21	&	New	\\
		16	&	3	&	0.75	&	93	&	H4		&	0.006	&	0.22	&	New	\\
		17	&	5	&	0.5	&	0	&	APR4	&	$<1\times 10^{-3}$	&	0.23	&	[1]	\\
		18	&	5	&	0.5	&	0	&	ALF2	&	0.01		&	0.27	&	[1]	\\
		19	&	5	&	0.5	&	0	&	H4		&	0.02		&	0.26	&	[1]	\\
		20	&	5	&	0.5	&	0	&	MS1		&	0.05		&	0.27	&	[1]	\\
		21	&	5	&	0.75	&	0	&	APR4	&	0.008	&	0.25	&	[1]	\\
		22	&	5	&	0.75	&	0	&	ALF2	&	0.05		&	0.28	&	[1]	\\
		23	&	5	&	0.75	&	0	&	H4		&	0.05		&	0.27	&	[1]	\\
		24	&	5	&	0.75	&	0	&	MS1		&	0.08		&	0.28	&	[1]	\\
		25	&	5	&	0.75	&	33	&	APR4	&	0.005	&	0.30	&	[2]	\\
		26	&	5	&	0.75	&	33	&	ALF2	&	0.03		&	0.27	&	[2]	\\
		27	&	5	&	0.75	&	33	&	H4		&	0.04		&	0.27	&	[2]	\\
		28	&	5	&	0.75	&	32	&	MS1		&	0.07		&	0.28	&	[2]	\\
		29	&	5	&	0.75	&	63	&	APR4	&	0.001	&	0.27	&	[2]	\\
		30	&	5	&	0.75	&	63	&	ALF2	&	0.007	&	0.28	&	[2]	\\
		31	&	5	&	0.75	&	63	&	H4		&	0.01		&	0.25	&	[2]	\\
		32	&	5	&	0.75	&	63	&	MS1		&	0.01		&	0.27	&	[2]	\\
		33	&	5	&	0.75	&	94	&	APR4	&	$<1\times 10^{-3}$	&	0.24	&	[2]	\\
		34	&	5	&	0.75	&	94	&	ALF2	&	$<1\times 10^{-3}$	&	0.26	&	[2]	\\
		35	&	5	&	0.75	&	94	&	H4		&	0.001	&	0.28	&	[2]	\\
		36	&	5	&	0.75	&	93	&	MS1		&	0.01		&	0.27	&	[2]	\\
		37	&	7	&	0.5	&	0	&	APR4	&	$<1\times 10^{-3}$	&	0.23	&	[1]	\\
		38	&	7	&	0.5	&	0	&	ALF2	&	$<1\times 10^{-3}$	&	0.27	&	[1]	\\
		39	&	7	&	0.5	&	0	&	H4		&	0.003	&	0.29	&	[1]	\\
		40	&	7	&	0.5	&	0	&	MS1		&	0.02		&	0.30	&	[1]	\\
		41	&	7	&	0.75	&	0	&	APR4	&	$<1\times 10^{-3}$	&	0.27	&	[1]	\\
		42	&	7	&	0.75	&	0	&	ALF2	&	0.02		&	0.29	&	[1]	\\
		43	&	7	&	0.75	&	0	&	H4		&	0.04		&	0.29	&	[1]	\\
		44	&	7	&	0.75	&	0	&	MS1		&	0.07		&	0.30	&	[1]	\\
		45	&	7	&	0.75	&	33	&	H4		&	0.03		&	0.28	&	New	\\\hline
	\end{tabular}
	\end{center}
\end{table*}

\begin{table*}
\begin{center}
	\caption{The bolometric corrections for {\it ugrizJHK}--band magnitude calculated from the result of the model APR4Q3a75 in~\cite{2014ApJ...780...31T}.}
	\label{tb:bc}
	\begin{tabular}{c|ccccccccc}\hline
			Rescaled Time&\multicolumn{8}{c}{Bolometric Correction $\Delta M_{\it X} ~[{\rm mag}]$ ($X$: band filter)}\\
			$\displaystyle\left(\frac{t}{{\rm day}}\right)\left(\frac{0.01~M_\odot}{M_{\rm ej}}\right)^{1/3.2}$	&	$X=$&{\it u}	&	{\it g}	&	{\it r}		&	{\it i}		&	{\it z}	&	{\it J}	&	{\it H}		&	{\it K}\\\hline\hline
   1.5&&  -0.28&  -0.45&  -0.47&  -0.71&  -0.97&  -1.61&  -2.37&  -4.55\\
   2.0&&  -0.19&  -0.57&  -0.59&  -0.32&  -0.48&  -0.07&   1.18&   0.41\\
   2.5&&   0.25&  -0.13&  -0.20&   0.05&  -0.28&  -1.50&   0.86&   2.22\\
   3.0&&   0.93&   0.93&   0.60&   0.56&   0.37&  -1.98&  -4.29&  -3.78\\
   3.5&&   0.47&   1.25&   1.36&   1.26&   1.04&  -1.38&  -4.65&  -6.36\\
   4.0&&  -0.34&   1.29&   1.20&   1.60&   1.44&  -0.74&  -3.31&  -6.12\\
   4.5&&   0.03&   0.74&   1.15&   1.51&   1.65&  -0.29&  -2.33&  -5.25\\
   5.0&&  -0.39&   0.37&   0.89&   0.99&   1.76&   0.42&  -1.73&  -4.10\\
   5.5&&  -0.69&   0.68&   0.74&   0.44&   1.13&   0.94&  -0.69&  -3.57\\
   6.0&&  -1.21&   0.41&   0.74&   0.50&   1.15&   1.05&  -0.66&  -3.55\\
   6.5&&  -3.65&  -0.58&   1.05&   0.94&   1.28&   1.66&  -0.67&  -3.76\\
   7.0&&  -4.40&  -0.87&  -0.16&   1.04&   1.82&   1.46&  -0.07&  -3.68\\
   7.5&&  -1.74&  -2.55&   0.14&   1.30&   2.09&   1.39&   0.01&  -3.24\\
   8.0&&  -1.36&  -2.87&   0.15&   1.31&   2.11&   1.38&   0.04&  -3.11\\
   8.5&&  -3.75&  -1.90&  -0.35&   1.16&   2.26&   1.53&   0.15&  -2.40\\
   9.0&&  -4.70&  -1.57&  -0.57&   1.08&   2.29&   1.59&   0.21&  -2.10\\
   9.5&&  -3.93&  -0.36&   0.62&   1.24&   2.50&   2.31&   0.31&  -1.34\\
  10.0&&  -4.00&  -0.44&   0.56&   1.20&   2.47&   2.29&   0.32&  -1.30\\
  10.5&&   3.18&  -0.52&   0.84&   1.34&   2.74&   2.43&   0.30&  -1.11\\
  11.0&&  --&  -0.60&   1.97&   1.92&   3.79&   2.93&   0.20&  -0.54\\
  11.5&&  --&  -0.68&   1.89&   1.87&   3.73&   2.90&   0.21&  -0.51\\
  12.0&&  --&   0.29&   1.59&   2.14&   3.74&   2.93&   0.43&   0.33\\
  12.5&&  --&   0.53&   1.44&   2.18&   3.70&   2.91&   0.50&   0.61\\
  13.0&&  --&   0.43&   1.36&   2.12&   3.64&   2.89&   0.51&   0.63\\
  13.5&&  --&  --&   2.05&   3.59&   3.00&   3.00&   1.27&   1.04\\
  14.0&&  --&  --&   2.19&   3.97&   2.77&   3.02&   1.49&   1.17\\\hline
	\end{tabular}
	\end{center}
\end{table*}

\bibliographystyle{apj}
%%\begin{thebibliography}{108}
%%\expandafter\ifx\csname natexlab\endcsname\relax\def\natexlab#1{#1}\fi

%%\end{thebibliography}
\bibliography{ref.bib}

\end{document}